\documentclass{amsart}

\usepackage[english]{babel}
\usepackage[utf8]{inputenc}
\usepackage{physics}
\usepackage{graphicx}
\usepackage[colorinlistoftodos]{todonotes}
\usepackage{url}
\usepackage{multicol}
\usepackage{subcaption}
\usepackage{graphicx}
\usepackage{textcomp}
\usepackage{xcolor}
\usepackage{enumerate}
\usepackage{amsmath}
\usepackage{tikz}
\usepackage{tikz-3dplot}
\usepackage{amssymb}

\newtheorem{theorem}{Theorem}[section]

\theoremstyle{definition}
\newtheorem{definition}[theorem]{Definition}
\newtheorem{proposition}[theorem]{Proposition}
\newtheorem{question}[theorem]{Question}
\newtheorem{corollary}[theorem]{Corollary}
\newtheorem{example}[theorem]{Example}

\theoremstyle{remark}
\newtheorem{remark}[theorem]{Remark}

\numberwithin{equation}{section}

\begin{document}

\title{A Commuting Hamiltonian Framework for Quantum Time Transfer}

%    Information for first author
\author{Nicholas R. Allgood}
%    Address of record for the research reported here
\address{University of Maryland Baltimore County}
%    Current address
\curraddr{1000 Hilltop Circle Baltimore, MD 21250}
\email{allgood1@umbc.edu}

\date{\today}


\begin{abstract}
%We develop a mathematical framework for quantum time transfer based on commuting families of Hamiltonians and synchronization observables. By encoding time information into the spectral decomposition of local clock observables, we formalize the preservation of timing correlations across composite quantum systems evolving under compatible dynamics. Using operator algebraic techniques, we characterize synchronization subspaces, establish stability under commuting perturbations, and classify compatible Hamiltonians via group-theoretic methods. The framework accommodates degenerate spectra and enables a perturbative analysis of long-time drift. While motivated by quantum synchronization and time transfer, the results provide a general algebraic structure for preserving relational observables in multipartite quantum dynamics. Concrete examples and potential extensions to categorical and infinite-dimensional formulations are discussed.

We develop a mathematical framework for quantum time transfer based on
commuting families of Hamiltonians and synchronization observables. The
synchronization subspace is defined as the kernel of a difference operator
between local clocks, and we show that this subspace is preserved exactly by a
commutative $*$-subalgebra of Hamiltonians compatible with the clocks. Our first
main result establishes \emph{perturbative stability}: for
$\epsilon$-compatible dynamics, where the commutator with the synchronization
operator is bounded in norm by $\epsilon$, we prove quantitative drift bounds
showing that timing correlations degrade at most linearly in time with slope
proportional to $\epsilon$. Our second main result provides a
\emph{representation-theoretic classification}: in the presence of a finite group
symmetry, the synchronization subspace coincides with the diagonal isotypic
component in the tensor product decomposition, and synchronization preservation
is characterized by the commutant algebra of the group action. These results
identify synchronization as a structural invariant of operator algebras,
connecting approximate commutation, kernel-preserving dynamics, and symmetry
protection. Beyond quantum time transfer, the framework suggests categorical and
resource-theoretic generalizations and contributes to the broader study of
operator-algebraic invariants in multipartite quantum dynamics.
\end{abstract}

\maketitle

\section{Introduction}

Synchronization of quantum systems across distributed networks is a fundamental
problem in quantum information and metrology. Operational protocols based on
entangled photons, shared oscillators, and classical post-processing have been
demonstrated experimentally \cite{ho2013clocksync, lafler2024twoway,
PhysRevApplied.20.024064}, but these methods lack a general algebraic framework
for analyzing when timing correlations are structurally preserved under quantum
evolution.

This paper develops such a framework in terms of operator algebras and
representation theory. The central objects are commuting families of Hamiltonians
and clock observables whose spectra encode discrete time labels. We show that the
synchronization subspace, defined as the kernel of a difference operator, is
preserved precisely by a commutative $*$-subalgebra of Hamiltonians compatible
with the clocks. This yields a classification theorem for synchronization-preserving
dynamics in terms of the commutant structure of the observables.

Two main results distinguish our approach:

\begin{itemize}
\item[(1)] \textbf{Perturbative stability.} We introduce the notion of
$\epsilon$-compatible dynamics, in which the Hamiltonian nearly commutes with the
synchronization operator. We prove quantitative drift bounds showing that timing
correlations degrade at most linearly in time with slope proportional to
$\epsilon$. This establishes the first perturbative stability result for
synchronization in quantum dynamics.

\item[(2)] \textbf{Representation-theoretic classification.} Using finite group
representation theory, we show that synchronization corresponds to spectral
alignment of irreducible representations across subsystems. The synchronized
subspace is identified with the diagonal isotypic component, and synchronization
preservation is characterized by a symmetry-respecting commutant algebra. This
reveals synchronization as a structural invariant of operator algebras and group
actions, independent of physical implementation.
\end{itemize}

These results elevate quantum time transfer from an operational protocol to a
general mathematical theory of relational observables. Beyond their original
motivation, they contribute to the study of kernel-preserving dynamics, approximate
commutation, and symmetry-protected subspaces in operator algebras. They also
suggest categorical generalizations, where synchronization-preserving unitaries are
morphisms in a category of observables with compatible dynamics. Such structures
connect naturally to broader themes in operator algebras, quantum information, and
mathematical physics.

\section{Preliminaries}

Quantum time transfer protocols traditionally rely on entangled photon pairs distributed between distant parties, with synchronization achieved through time-tagging and classical post-processing \cite{ho2013clocksync, lafler2024twoway, PhysRevApplied.20.024064}. These approaches focus on operational accuracy but do not model timing information as an intrinsic feature of quantum observables or dynamics. Recent work on relational quantum observables \cite{rovelli1996relational} and quantum reference frames \cite{bartlett2007reference} explores related ideas, but typically emphasizes full reference frame alignment rather than timing synchronization per se. In contrast, our framework treats synchronization as a structural property of specific commuting observables and characterizes the dynamical preservation of timing correlations at the algebraic level. This approach provides a general mathematical theory complementary to operational methods, capable of analyzing synchronization stability and symmetry structures beyond particular physical implementations.

We work throughout with finite-dimensional Hilbert spaces and bounded linear operators. Let $\mathcal{H}_A$ and $\mathcal{H}_B$ denote two such spaces and let $\mathcal{H} := \mathcal{H}_A \otimes \mathcal{H}_B$ denote their tensor product. We denote by $\mathcal{B(H)}$ the algebra of bounded linear operators on $\mathcal{H}$, which coincides with the set of all linear maps in finite dimensions. Let $\ket{\Psi}_{AB} \in \mathcal{H}$ denote a general bipartite quantum state which may be entangled or separable. When $\ket{\Psi|}_{AB}$ happens to be separable, we write $\ket{\Psi}_{AB} = \ket{\psi}_A \otimes \ket{\phi}_B$ with $\ket{\psi}_A \in \mathcal{H}_A$, $\ket{\phi}_B \in \mathcal{H}_B$. In this work, a \textit{clock} is modeled as a self-adjoint observable whose eigenvalues represent time labels associated with discrete measurement outcomes.

\begin{definition} (Bounded Operators and Observables)
    An operator $A \in \mathcal{B(H)}$ is called self-adjoint if $A = A^{\dagger}$, where $A^{\dagger}$ denotes the Hermitian adjoint. A quantum observable is a self-adjoint operator. In finite-dimensional Hilbert spaces, all linear operators are bounded. 
\end{definition} 

\begin{definition}[Range of an Operator]
Let \( A \in \mathcal{B}(\mathcal{H}) \) be a bounded linear operator on a Hilbert space \( \mathcal{H} \).  
The \textit{range} of \( A \), denoted \( \operatorname{ran}(A) \), is the set
\[
\operatorname{ran}(A) := \{ Av \mid v \in \mathcal{H} \} \subseteq \mathcal{H}.
\]
It consists of all vectors that can be expressed as \( Av \) for some \( v \in \mathcal{H} \).
\end{definition}

\begin{definition}[Tensor Product of Operators]
Let $A \in \mathcal{B}(\mathcal{H}_A)$ and $B \in \mathcal{B}(\mathcal{H}_B)$. 
The tensor product operator $A \otimes B \in \mathcal{B}(\mathcal{H}_A \otimes \mathcal{H}_B)$ is defined on simple product tensors by
\begin{equation}
    (A \otimes B)(\,|\psi\rangle_A \otimes |\phi\rangle_B) 
    = (A|\psi\rangle_A) \otimes (B|\phi\rangle_B),
\end{equation}
for $|\psi\rangle_A \in \mathcal{H}_A$, $|\phi\rangle_B \in \mathcal{H}_B$,
and extends by linearity to all states 
$|\Psi\rangle_{AB} \in \mathcal{H}_A \otimes \mathcal{H}_B$.
\end{definition}

\begin{definition} (Unitary Time Evolution)
    Let $H \in \mathcal{B(H)}$ be self-adjoint. The unitary evolution generated by $H$ is \
    \begin{equation}
        U(t) := e^{-iHt},\ t \in \mathbb{R},
    \end{equation}
    which defines a one-parameter unitary group satisfying \cite{reed_simon}:
    \begin{gather*}
        U(0) = I, \\
        U(t+s) = U(t)U(s), \\ 
        \frac{d}{dt}U(t) = -iHU(t) 
    \end{gather*}

\end{definition} 

\begin{definition}[Commuting Local Hamiltonians]
Let \( \mathcal{H}_A, \mathcal{H}_B \) be finite-dimensional Hilbert spaces, and let
\[
H := H_A \otimes I + I \otimes H_B
\]
for self-adjoint operators \( H_A \in \mathcal{B}(\mathcal{H}_A) \), \( H_B \in \mathcal{B}(\mathcal{H}_B) \).

We say the system evolves under a \emph{commuting bipartite Hamiltonian} when this decomposition satisfies
\[
[H_A \otimes I, I \otimes H_B] = 0.
\]

This commutation follows trivially from the disjoint supports of the tensor factors, but we emphasize that commutativity is a property of the decomposition structure of \( H \), not of the operator \( H \) alone.
\end{definition}

\begin{definition} (Commuting Hamiltonian System)
  Let $H_A \in \mathcal{B}(\mathcal{H}_A), H_B \in \mathcal{B}(\mathcal{H}_B)$ be self-adjoint. Define the total Hamiltonian as joint Hamiltonian on $\mathcal{H}_A \otimes \mathcal{H}_B$ is:
    \begin{equation}
        H := H_A \otimes I + I \otimes H_B \in \mathcal{B(H)}.
    \end{equation}
    We say this is a commuting Hamiltonian system since 
    \begin{equation}
        [H_A \otimes I, I \otimes H_B] = 0
    \end{equation}
    Let $H_A \in \mathcal{B}(\mathcal{H}_A), H_B \in \mathcal{B}(\mathcal{H}_B)$ be self-adjoint observables interpreted as local clocks. These observables may be diagonal in a fixed basis, encoding time-tagged detection events. Suppose:
    \begin{equation}
        [T_A, H_A] = 0,\ [T_B, H_B] = 0.
    \end{equation}
   This ensures that the clock observable is invariant under local evolution where its eigenspaces are preserved over time.

\end{definition}

\begin{definition}[Clock Observables]
Let \( H_A, H_B \) be local Hamiltonians on \( \mathcal{H}_A \), \( \mathcal{H}_B \), respectively.  
We say that \( T_A \in \mathcal{B}(\mathcal{H}_A) \) and \( T_B \in \mathcal{B}(\mathcal{H}_B) \) are \textit{clock observables} if:

\begin{itemize}
  \item They are self-adjoint,
  \item They commute with the local Hamiltonians: \( [T_A, H_A] = 0 \), \( [T_B, H_B] = 0 \),
\end{itemize}

so that each pair \( \{T_A, H_A\} \), \( \{T_B, H_B\} \) admits a common orthonormal eigenbasis. We interpret the eigenvalues of \( T_A \), \( T_B \) as local time labels associated to the shared eigenbasis with the local dynamics.
\end{definition}

\begin{proposition}
    If $[T_A, H_A] = 0$ and $[T_B, H_B] = 0$, then $\mathcal{A}$ is a commutative *-subalgebra of $\mathcal{B(H)}$. 

    \begin{proof}
        Each generator acts on a disjoint subsystem. The assumed commutativity ensures all pairwise commutators vanish. Since all generators are self-adjoint and include the identity, $\mathcal{A}$ is closed under adjoints and unital, hence a commutative *-subalgebra.
    \end{proof}
\end{proposition}

\begin{remark} 
    While the Hamiltonian $H = H_A \otimes I + I \otimes H_B$ is separable, it can evolve initial entangled states without destroying timing correlations, provided the compatibility conditions $[T_x,H_x] = 0$ hold for $x = \{A, B\}$. Since $H = H_A \otimes I + I \otimes H_B$ is a sum of local terms, it does not itself generate entanglement from a product state. However, if the initial state is already entangled, the compatibility conditions $[T_x,H_x] = 0$ for $x = \{A,B\}$ ensure that timing correlations encoded in the joint eigenspaces of $T_A$ and $T_B$ are preserved under the evolution.
\end{remark}

%2.8

\begin{theorem}[Preservation of Synchronization Under Compatible Dynamics]

    Let $\mathcal{H}_A$ and $\mathcal{H}_B$ be finite-dimensional Hilbert spaces. Let $T_A \in \mathcal{B(H}_A)$ and $T_B \in \mathcal{B(H}_B)$ be self-adjoint operators that are local time observables. Let $H_A \in \mathcal{B(H}_A)$ and $H_B \in \mathcal{B(H}_B)$ be self-adjoint Hamiltonians satisfying:
    \begin{align}
        [T_A, H_A] = 0,\ [T_B,H_B] = 0.
    \end{align}

    Define the global Hamiltonian
    \begin{equation}
        H := H_A \otimes I + I \otimes H_B
    \end{equation}

    and the global unitary evolution 
    \begin{equation}
        U(t) := e^{-iHt}
    \end{equation}

    Let 
    \begin{equation}
        K := T_A \otimes I - I \otimes T_B
    \end{equation}
    and define the synchronized subspace
    
    \begin{equation}
        \mathcal{K} := \ker(K) \subseteq \mathcal{H}_A \otimes \mathcal{H_B}.
    \end{equation}

    Then $[K,H] = 0$ for all $t \in \mathbb{R}$ implying $U(t)\mathcal{K} \subseteq \mathcal{K}$ is the synchronization subspace preserved under time evolution.

    \begin{proof}

        Since $[T_A,H_A] = 0$ and $[T_B,H_B] = 0$, it follows that:
        
        \begin{align}
            [T_A \otimes I, H_A \otimes I] &= 0, & [I \otimes T_B, I \otimes H_B] &= 0, \\
            [T_A \otimes I, I \otimes H_B] &= 0, & [I \otimes T_B, H_A \otimes I] &= 0 .
        \end{align}

        The operators act non-trivially on disjoint tensor factors thus:
        
        \begin{multline}
            [K,H] = [T_A \otimes I - I \otimes T_B, H_A \otimes I + I \otimes H_B] = [T_A \otimes I, H_A \otimes I] \\
            + [T_A \otimes I, I \otimes H_B] - [I \otimes T_B, H_A \otimes I] - [I\otimes T_B, I \otimes H_B] = 0
        \end{multline}
        Therefore, $K$ commutes with $H$. Now, for any $\ket{\psi(0)} \in \mathcal{K}$, we have 
        \begin{equation}
            K\ket{\psi(0)} = 0.
        \end{equation}

        Since $[K,U(t)] =  0$ by functional calculus, due to $K$ and $H$ commuting, we have:
        \begin{equation}
            KU(t)\ket{\psi(0)} = U(t)K\ket{\psi(0)} = U(t)(0) = 0.
        \end{equation}
        Thus we can state 
        \begin{equation}
            U(t)\ket{\psi(0)} \in \mathcal{K},
        \end{equation}
        which shows the synchronization subspace $\mathcal{K}$ is preserved under the evolution $U(t)$ for all $t \in \mathbb{R}$.
        
    \end{proof}

\end{theorem}

We interpret \( T_A \) and \( T_B \) as local clock observables for subsystems A and B, respectively.  They model internal time labels such as arrival-time indices or timing tags that are associated with a shared basis in which the system evolves. These observables are assumed to commute with the local Hamiltonians so that time information encoded in the eigenstructure remains stable under evolution.  In quantum optics protocols, such observables can correspond to coarse-grained time bins or arrival-time registers at spatially separated nodes. Given a clock that is atomic or pulsed-laser, the time bins could also be defined as the pulse duration.

\begin{remark}
    We emphasize that our framework models time as an internal label encoded in the spectral structure of a fixed observable. To clarify, spectral structure in this instance refers to the decomposition of $\mathcal{H}$ into orthogonal eigenspaces of the operator together with the associated eigenvalues. This approach differs from dynamical models in which time and energy are conjugate variables and may not commute. Here, the condition $[T,H] = 0$ ensures the observables eigenspaces are preserved under evolution. This is appropriate in scenarios where timing labels correspond to fixed arrival-time registers or bins with evolution corresponding to local decoherence-free subspaces.
\end{remark}

\begin{remark}
While the present framework is developed in finite-dimensional Hilbert spaces, many of the core constructions, such as synchronization operators, commuting observables, and kernel subspaces, admit natural generalizations to the infinite-dimensional setting. However, care must be taken with domain issues, unbounded operators, and spectral theory subtleties that arise when working with infinite-dimensional time observables (e.g., in continuous-variable systems or clock states modeled by Fock space). We leave such generalizations for future work, noting that finite dimensions suffice for the discrete synchronization protocols studied here.
\end{remark}

\begin{question}
    Under what conditions does the evolution  $U(t)$ preserve timing correlations for states that do not lie exactly in the kernel $\ker(T_A \otimes I - I \otimes T_B)$? Can these structures be robustly maintained under approximate commutation or small perturbations?
\end{question}

The subsequent sections address these questions by identifying classes of compatible and \(\epsilon\)-compatible Hamiltonians, and developing an algebraic and representation-theoretic framework for synchronization preservation.

\section{Time Observables and Synchronization Subspaces}

We now formalize the concept of a \textit{clock observable} and define the synchronization structure that arises from spectral alignment in bipartite quantum systems. Our framework encodes timing information into the eigenstruture of a self-adjoint operator, enabling unitary evolution to preserve correlations between time-labeled subsystems.

\subsection{Time Observables and Spectral Encoding}

Let \( \mathcal{H} \) be a finite-dimensional Hilbert space.  
A \textit{time observable} is a self-adjoint operator \( T \in \mathcal{B}(\mathcal{H}) \) with a discrete spectrum,  
which encodes timing information via its eigenvalues.  
We interpret the spectrum \( \{t_j\} \subset \mathbb{R} \) as \textit{time labels},  
and the associated eigenvectors \( \{\ket{j}\} \) as \textit{clock states}.

If \( T \) is compatible with a Hamiltonian \( H \), in the sense that \( [T, H] = 0 \),  then \( T \) and \( H \) are simultaneously diagnolizable,  and the clock structure is preserved under time evolution. In a fixed eigenbasis of \( T \), we may write:
\[
T = \sum_{j=0}^{d-1} t_j \ket{j}\bra{j},
\]
where each \( t_j \in \mathbb{R} \) is a time label. When $T$ and $H$ are simultaneously diagnolizable, they share a complete eigenbasis. In physical terms this means that the masurement outcomes of the clock observable $T$ are constants of motion under the evolution granted by $H$. The system's state may change within each eigenspace of $T$, but no probability amplitude leaks between eigenspaces. This ensures the timing information encoded by $T$ remains stable throughout the entire evolution. 

%\subsection{Spectral Encoding of Time}

%Let $\mathcal{H} \cong \mathbb{C}^d$ be a finite-dimensional Hilbert space with orthonormal basis ${\ket{j}}_{j=0}^{d-1}$. A \textit{time observable} is a self-adjoint operator $T \in \mathcal{B(H)}$ of the form:

%\begin{equation}
 %   T = \sum_{j=0}^{d-1} t_j \ket{j}\bra{j},\ t_j \in \mathbb{R}
%\end{equation}

%The eigenvalues ${t_j}$ are interpreted as \textit{time tags} and the eigenvectors ${\ket{j}}$  as the corresponding \textit{clock states}. Upon measurement, the observable $T$ yields a value $t_j$ with a probability of $|\braket{j, \psi|}|^2$, projecting the system into the state $\ket{j}$. Time becomes an intrinsic attribute of the quantum system, embedded in its spectral structure \cite{reed_simon}. 

\subsection{Synchronization in Composite Systems}

Let $\mathcal{H}_A \cong \mathbb{C}^d$ and $\mathcal{H}_B \cong \mathbb{C}^d$ be Hilbert spaces for two quantum subsystems. Define identical local time observables:

\begin{equation}
    T_X = \sum_{j=0}^{d-1} t_j \ket{j}\bra{j},\ X \in \{A,B\}
\end{equation}
with shared time labels ${t_j}$. A bipartite state $\ket{\psi} \in \mathcal{H}_A \otimes \mathcal{H}_B$ is said to be synchronized if the measurement outcomes of $T_A \otimes I$ and $I \otimes T_B$ are perfectly correlated if they yield the same time label with probability one. The subspace of all such synchronized states is given by the kernel:

\begin{equation}
  \mathcal{K} :=  \ker(T_A \otimes I - I \otimes T_B)
\end{equation}

This subspace plays a central role in our framework. In subsequent sections, we will characterize the class of dynamics that preserve $\mathcal{K}$ and study its structural and perturbative stability. 

\begin{remark}
    This abstraction captures a core idea in quantum optics: the use of entangled photon pairs and time-of-arrival tagging to synchronize remote clocks. In contrast, our formulation encodes timing internally within the observables themselves, enabling synchronization to persist under unitary evolution. This operator-theoretic shift allows us to model timing correlations intrinsically without relying on external reference clocks or post-processing.
\end{remark}

\section{Structural Consequences of Commuting Dynamics}

We examine the structural properties of time evolution under commuting Hamiltonians, with a focus on the preservation of timing correlations. Building on Theorem 2.8, we demonstrate that the subspace of bipartite states exhibiting perfect synchronization is invariant under unitary evolution when the dynamics are compatible with the observables. Specifically, we show that the subspace $\mathcal{K} := \ker(T_A \otimes I - I \otimes T_B)$ is preserved by all Hamiltonians of the form $H_A \otimes I + I \otimes H_B$ that commute with the corresponding observables. 

\begin{definition}(Kernel of an Operator)
    Let $A \in \mathcal{B}(\mathcal{H})$ be the kernel or null space of $A$ is defined as:
    \begin{equation*}
        \ker(A) := \{ {\ket{\psi} \in \mathcal{H} \mid A\ket{\psi} = 0 \}}
    \end{equation*}
    We say that a vector $\ket{\psi}$ is in the kernel of $A$ if it is annihilated by $A$. If $A$ is self-adjoint, the kernel is a closed subspace of $\mathcal{H}$. 

    We focus on the operator $T_A \otimes I - I \otimes T_B$ and interpret $\ker(T_A \otimes I - I \otimes T_B)$ as the set of all bipartite states whose local clock measurements are perfectly correlated. 
\end{definition}

%4.2 currently? 
\begin{theorem}[Spectral Stability and Synchronization under Compatible Evolution]
    Let $\mathcal{H}_A$ and $\mathcal{H}_B$ be finite-dimensional Hilbert spaces. Let $T_A \in \mathcal{B(H}_A)$ and $T_B \in \mathcal{B(H}_B)$ be self-adjoint observables and let $H_A \in \mathcal{B(H}_A)$ and $H_B \in \mathcal{B(H}_B)$ be self-adjoint Hamiltonians satisfying:
    \begin{align}
        [T_A,H_A] =0,\  [T_B,H_B] = 0.
    \end{align}

    Define the global Hamiltonian:
    \begin{equation}
        H := H_A \otimes I + I \otimes H_B,
    \end{equation}
    
    the global unitary evolution:
    \begin{equation}
        U(t) := e^{-iHt}
    \end{equation}

    the synchronization operator:
    \begin{equation}
        K := T_A \otimes I - I \otimes T_B
    \end{equation}
    
    with the synchronized subspace
    
    \begin{equation}
        \mathcal{K} := \ker(K).
    \end{equation}

    Then the observables $T_A$ and $H_A$ admit a common eigenbasis; similarly for $T_B$ and $H_B$. The synchronization operator $K$ is diagonal in the joint eigenbasis. The global unitary evolution $U(t)$ preserves the eigenspaces of $K$ and thus $U(t)\mathcal{K} \subseteq \mathcal{K}$. The spectra of the time observables $T_A$ and $T_B$ are preserved under time evolution. Synchronization is encoded through matching time eigenvalues that are spectrally stable under compatible dynamics \cite{kadison_ringrose, reed_simon}. 

    \begin{proof}
    Since $[T_A, H_A] = 0$ and $[T_B,H_B] = 0$, standard results on commuting self-adjoint operators imply that:
    \begin{enumerate}
        \item $T_A$ and $H_A$ can be simultaneously diagonalized by an orthonormal eigenbasis $\{\ket{j}\} \subset \mathcal{H}_A$.
        \item $T_B$ and $H_B$ can be simultaneously diagonalized by an orthonormal eigenbasis $\{\ket{k}\} \subset \mathcal{H}_B$
    \end{enumerate}
    The global tensor product basis $\{\ket{J} \otimes \ket{K}\}$ simultaneously diagonalizes all four operators:
    \begin{align}
        T_A &\otimes I, & I &\otimes T_B, & H_A &\otimes I, & I &\otimes H_B.
    \end{align}

    In this basis, the synchronization operator $K = T_A \otimes I - I \otimes T_B$ is diagonal with the eigenvalues $t_j - t_k$. A state $\ket{\psi}$ belongs to $\mathcal{K}$ if and only if it is supported on the tensor products $\ket{j} \otimes \ket{j}$ where $t_j = t_k$. $H$ acts diagonally in this basis and the global evolution $U(t) = e^{-iHt}$ acts by multiplying each basis vector by a phase factor, preserving the support structure. If $\ket{\psi(0)} \in \mathcal{K}$, we have $\ket{\psi(t)} := U(t)\ket{\psi(0)} \in \mathcal{K}$ for all $t \in \mathbb{R}$. Because $[H,K] = 0$, it follows by functional calculus that $[U(t), K] = 0$, reinforcing that $\mathcal{K}$ is preserved under time volution. Finally, since $T_A \otimes I$ and $I \otimes T_B$ are diagonal in the same basis, their spectra are preserved under the action of $U(t)$.
 
    \end{proof}
\end{theorem}

\begin{remark}
This result shows that synchronization is dynamically preserved when each subsystem evolves under a Hamiltonian compatible with its local clock observable. In this framework, time is encoded in the spectral decomposition of \( T_A \) and \( T_B \), and preservation of the synchronization kernel corresponds to maintaining equality of time labels across the bipartite system.
\end{remark}

\begin{figure}[h]
\centering 
%\tdplotsetmaincoords{60}{120}%
%\tdplotsetrotatedcoords{0}{20}{0}%
\begin{tikzpicture}
\matrix [column sep=1cm, row sep=1cm]
{
\node(a) {$\mathcal{H}_A \otimes \mathcal{H}_B$}; &; &; &; &; \node(b) {$\mathcal{H}_A \otimes \mathcal{H}_B$}; \\

\node(c) {Projection $\Pi_\mathcal{K}$}; & & & &  \node(d) {Projection $\Pi_\mathcal{K}$}; \\

\node(e) {$\mathcal{K} = \ker(T_A \otimes I - I \otimes T_B)$}; & & & & \node(f) {$\mathcal{K}$};\\
};

\draw[->,thick] (a.east) -- (b.west) node[above, align=center,midway]{Unitary Evolution:\\$U(t)$};
\draw[->, thick] (a.south) -- (c.north) node[]{};
\draw[->, thick] (b.south) -- (d.north) node[]{};
\draw[->, thick] (c.south) -- (e.north) node[]{};
\draw[->, thick] (d.south) -- (f.north) node[]{};
\draw[->,thick] (e.east) -- (f.west) node[above, align=center,midway]{Evolution:\\$U(t)|_\mathcal{K}$};

\end{tikzpicture}
\caption{
Commutative diagram illustrating synchronization-preserving dynamics. When the global Hamiltonian \( H = H_A \otimes I + I \otimes H_B \) commutes with the synchronization operator \( T_A \otimes I - I \otimes T_B \), the evolution \( U(t) = e^{-iHt} \) preserves the synchronization subspace \( \mathcal{K} \). The diagram expresses that time evolution and projection onto \( \mathcal{K} \) commute, ensuring synchronized states remain invariant.
}

\end{figure}
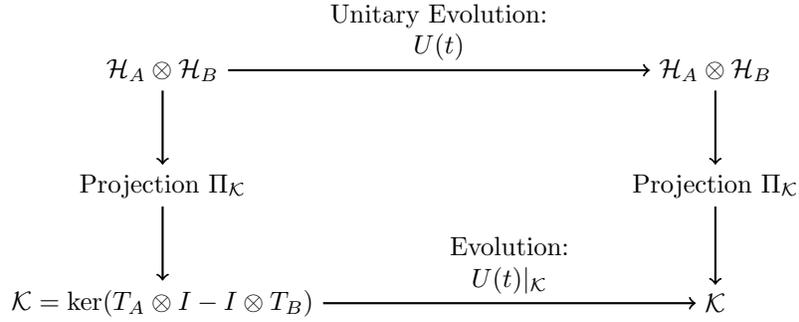

Having established the structural preservation of timing correlations under compatible dynamics, we now turn to explicit constructions of time observables and Hamiltonians. In particular, we introduce simple clock models where timing information is embedded directly into the spectral structure of self-adjoint operators, and characterize the allowable dynamics that preserve these encoded time tags.

\section{Constructing Compatible Clock Observables and Hamiltonians}

In this section we formalize the notion of \textit{clock observables} and characterize the class of Hamiltonians that preserve their eigenstructure. We begin by constructing explicit diagonal observables whose eigenvalues serve as time labels. We then classify all Hamiltonians that commute with a given clock observable, showing that such dynamics preserve the timing structure encoded into the eigenbasis of the observable. 

\subsection{Characterization of Compatible Hamiltonians}

\begin{definition}[Clock Observable]
Let $\mathcal{H}$ be a finite dimensional Hilbert space. A \textit{clock observable} is a self-adjoint operator $T \in \mathcal{B}(\mathcal{H})$ with non-degenerate spectrum $\sigma(T) = (t_o, t_1, \dots, t_{d-1}) \in \mathbb{R}$. We interpret each eigenvalue $t_j$ as a discrete time label. The eigenvectors of $T$ define a canonical orthonormal basis $\{\ket{j}\}$ in which $T = \sum t_j \ket{j} \bra{j}$. 

\end{definition}

\begin{remark}
    While Definition 5.1 introduces a clock observable that defines a canonical orthonormal basis, the concept of quantum coherence remains compatible with this framework. A quantum system described by such a clock observable can still exhibit coherence as a superposition of states within this or any other chosen basis. The existence of a self-adjoint clock observable with a non-degenerate spectrum and its associated orthonormal basis does not preclude the presence of quantum coherence. Coherence, defined by the superposition of quantum states within a specific basis, can readily exist alongside a precisely defined clock. Even when considering a clock observable with discrete time labels and a defined orthonormal basis, the principles of quantum coherence can still apply. The system can exist in a superposition of these 'time states,' representing a quantum description of time.
\end{remark}

\begin{theorem}
        Let $T \in \mathcal{B}(\mathcal{H})$ be a clock observable with the spectral decomposition of $T = \sum t_j \ket{j} \bra{j}$. Then every Hamiltonian $H \in \mathcal{B}(\mathcal{H})$ that commutes with $T$ satisfies $[H,T] = 0$ and is diagonal in the same basis. That is $H = \sum h_j \ket{j} \bra{j}$ for some real numbers $h_j \in \mathbb{R}$.

        \begin{proof}
            Since $T$ is self-adjoint with a non-generate spectrum, it's eigenspaces are one-dimensional. Therefore the commutant of $T$ consists of all operators that are diagonal in the same basis. If $[H,T] = 0$, then $H$ must preserve each eigenspace of $T$ and thus be diagonal in the eigenbasis of $T$. The self-adjoint-ness of $H$ ensures the coefficients $h_j$ are real.
        \end{proof}

\end{theorem}

\begin{corollary}[Algebra of Compatible Dynamics]
    The set of all Hamiltonians compatible with a clock observable $T$ forms a commutative unital *-subalgebra of $\mathcal{B}(\mathcal{H})$, given by $\mathcal{A} := \operatorname{alg}(T) = \{ f(T) \mid f(T): \sigma(T) \rightarrow \mathbb{R} \}$.

    This algebra describes the maximal set of dynamical generators that preserve the spectral structure and therefore the timing interpretations of the clock observables. 

    \begin{proof}
        Since $T$ is self-adjoint with a simple spectrum, the set of polynomials or bounded real functions applied to $T$ generates a commutative *-subalgebra of $\mathcal{B}(\mathcal{H}$. Every operator that commutes with $T$ must be diagonal in the eigenbasis of T and vice-versa. Hence $\operatorname{alg}(t)$ consists of all real-valued functions on the spectrum of $T$ and includes exactly those Hamiltonians that satisfy $[H,T] = 0$.
    \end{proof}
    
\end{corollary}

\begin{remark}
    This formation captures the minimal conditions under which timing structure encoded into the observable $T$ is preserved. If $T$ encodes discrete time labels, then any compatible Hamiltonian must respect those labels through its diagonal action. This constraint ensures that the timing information remains stable under time evolution and provides a mathematical foundation for physically meaningful synchronization dynamics.
\end{remark}

\begin{example}
    Let $\mathcal{H} = \mathbb{C}^3$ and define the clock observable $T:= \operatorname{diag}(0, 1, 2)$ in the standard basis $(\ket{0}, \ket{1}, \ket{2})$. The operator encodes discrete time labels with equally spaced values. The spectral decomposition is $T = 0 \cdot \ket{0} \bra{0} + 1 \cdot \ket{1} \bra{1} + 2 \cdot \ket{2} \bra{2}$. Any self-adjoint Hamiltonian $H \in \mathcal{B}(\mathbb{C}^3$ that satisfies $[H, T] = 0$ must be diagonal in this basis. For instance, $H_1 = \operatorname{diag}(1, 1, 1)$ generates trivial dynamics (global phase). $H_2 = \operatorname{diag}(\pi, -\pi, 0)$ defines nontrivial phase evolution that preserves timing structure. $H_3 = \operatorname{diag}(0, \sqrt(2), -1)$ also satisfies $[H_3, T] = 0$ and defines a valid time preserving Hamiltonian. However the Hamiltonian $H_4$: 
    \begin{equation*}
            H_4 = \begin{bmatrix}
                 0 & 1 & 0 \\
                1 & 0 & 0 \\
                 0 & 0 & 1 \\
        \end{bmatrix}
    \end{equation*}

    does not commute with $T$, since it includes off-diagonal terms in the eigenbasis of $T$. Therefore, it does not preserve the timing structure encoded in $T$ and $[H_4, T] \neq 0$.
    
    \end{example}

\begin{question}
    While our approach begins with clock observable $T$ and seeks compatible Hamiltonians $H$, one may consider the inverse of the problem: given a fixed Hamiltonian $H$, can one construct a non-trivial clock observable $T$ that commutes with $H$? This formulation reverses the roles of $T$ and $H$ and could lead to alternative classification results.
\end{question}

\subsection{Generalization to Arbitrary Dimensions}

\begin{proposition}
    Let $T \in \mathcal{B}(\mathcal{H})$ be a self-adjoint operator on a finite-dimensional Hilbert space $\mathcal{H}$ with a spectral decomposition:
    \begin{equation}
        T = \sum_{\lambda \in \sigma(T)} \lambda P_\lambda
    \end{equation}
    where each $P_\lambda$ is the orthogonal projection onto the eigenspace corresponding to the eigenvalue $\lambda$. Then any operator $H \in \mathcal{B}(\mathcal{H})$. that commutes with $T$ must satisfy
    \begin{equation}
        H = \sum_{\lambda \in \sigma(T)} P_\lambda H P_\lambda
    \end{equation}
    In particular, the set of all such operator forms a unital *-subalgebra of $\mathcal{B}(\mathcal{H})$, isomorphic to the direct sum:
    \begin{equation}
        \mathcal{A}_T := \bigoplus_{\lambda \in \sigma(T)} \mathcal{B}(\operatorname{ran}(P_\lambda))
    \end{equation}

    \begin{proof}
        Since $T$ is self-adjoint with a spectral decomposition $T = \sum \lambda P_\lambda $, it's commutant $\mathcal{A}_T := \{ H \in \mathcal{B}(\mathcal{H}) \mid [H, T] = 0\}$ consists of all operators that preserve the eigenspaces of $T$.  This implifes that $H$ must be block-diagonal with respect to the decomposition $\mathcal{H} = \bigoplus \operatorname{ran}(P_\lambda)$, i.e: 
    \begin{equation}
        H = \sum_{\lambda \in \sigma(T)} P_\lambda H P_\lambda
    \end{equation} with no cross terms between distinct eigenspaces. Each $P_\lambda H P_\lambda \in \mathcal{B}(\operatorname{ran}(P_\lambda))$, so the commutant algebra $\mathcal{A}_T$ is isomorphic to the direct sum of full matrix algebras:
    \begin{equation}
        \mathcal{A}_T \cong \bigoplus_{\lambda \in \sigma(T)} \mathcal{B}(C^{d_\lambda})
    \end{equation}
    where $d^\lambda = \operatorname{dim}(\operatorname{ran}(P_\lambda))$. This algebra is closed under adjoint multiplication, hence a unital *-subalgebra.

    \end{proof}
    
\end{proposition}

\begin{remark}
    The results above show that the structure of clock-compatible dynamics is entirely governed by the spectral properties of the clock observable. When the spectrum is non-degenerate, compatible Hamiltonians are strictly diagonal and when degeneracies are present, the dynamics decompose into independent unitary evolutions on each eigenspace. This characterization not only captures a wide class of timing-preserving dynamics, but also serves as a foundation for understanding how slight deviations from exact compatibility, such as those arising in experimental or noisy systems, affect the preservation of timing information.
\end{remark}

\section{Stability and Perturbations}

In physical implementations of quantum time transfer protocols, exact commutation between Hamiltonians and time observables is an idealization. In practice, small deviations arise due to experimental imperfections, noise, or environmental couplings. In this section we introduce a perturbative framework based on $\epsilon$-commutation and quantify the extent to which timing correlations are preserved under such approximately compatible dynamics. We focus on the operator $K := T_A \otimes I - I \otimes T_B$ which encodes synchronization through it's kernel. When the commutator $[H,K]$ is small in operator norm, where $||[H,K]|| \ll 1$, the resulting dynamics can be regarded as an approximately compatible evolution. In this regime, one can quantify the drift from perfect synchronization through perturbation theory \cite{Shankar}.

When $[H,K]=0$ but remains small in norm, we seek to understand how much deviation from $\ker(K)$ accumulates over time \cite{ceccherini2010representation}.  

\begin{definition}[$\epsilon$-Compatible Dynamics]
    Let $T_A \in \mathcal{B}(\mathcal{H}_A)$ and  $T_B \in \mathcal{B}(\mathcal{H}_B)$ and define the synchronization operator:
    \begin{equation}
        K := T_A \otimes I - I \otimes T_B.
    \end{equation}
    A self-adjoint Hamiltonian $H \in \mathcal{B}(\mathcal{H}_A \otimes  \mathcal{H}_B)$ is said to be $\epsilon$-compatible with the time observables if $||[H,K]|| \leq \epsilon$ for some $\epsilon \geq 0$. This condition quantifies the degree to which the Hamiltonian fails to exactly preserve the kernel of $K$ with $\epsilon = 0$, which corresponds to the exact commutation and perfect timing preservation. 

\end{definition}

 Similar commutator-based stability conditions have been studied in various contexts \cite{bravyi2006lieb, ceccherini2010representation, hastings2004lsm} to quantify approximate symmetries or conservation laws. We refer to such dynamics as $\epsilon$-compatible with the synchronization structure 

\begin{theorem}[Perturbative Kernel Drift under $\epsilon$-Compatible Dynamics]
    Let $H$ be a self-adjoint Hamiltonian on $\mathcal{H}_A \otimes \mathcal{H}_B$ and define $K:= T_A \otimes I - I \otimes T_B$ with a synchronization subspace $\mathcal{K} := \ker(K)$. Suppose $||[H,K]|| \leq \epsilon$ for some $\epsilon \geq 0$. Then for any normalized initial state $\ket{\psi(0)} \in \mathcal{K}$, the evolved state $\ket{\psi(t)} := e^{(-iHt)}\ket{\psi(0)}$ satisfies: 
    \begin{equation}
        ||K\ket{\psi(t)}|| \leq \epsilon |t|.
    \end{equation}

    where $|t|$ denotes the magnitude of evolution time. 
    
    \begin{proof}
        Let $\ket{\psi(t)} = U(t)\ket{\psi(0)}$, where $U(t) := e^{(-iHt)}$ and $\phi(t) := K\ket{\psi(t)}$. Then:
        \begin{align*}
            \frac{d}{dt}\varphi(t) = \frac{d}{dt}(Ke^{-iHT}\ket{\psi(0)} = -iKH|\ket{\psi(t)} = \\-iHK\ket{\psi(t)} + i[H,K] \ket{\psi(t)}.
        \end{align*}

        Using the fact that $\ket{\psi(0)} \in \ker(K)$ implies $\phi(0) = 0$, we solve this linear inhomogeneous equation with Duhamel's formula \cite{Duahmel}:
        \begin{equation}
            \varphi(t) = i\int_0^t e^{-iH(t-s)} [H,K]\ket{\psi(s)}ds.
        \end{equation}

        Taking the norms and applying the unitarity of $e^{(-iH(t-s) )}$ gives: 
        \begin{equation}
            ||\varphi(t)|| \leq \int_0^t ||[H,K]|| \cdot ||\psi(s)||ds \leq \epsilon|t|.
        \end{equation}
        Thus $||K\ket{\psi(t)} \leq \epsilon |t|$ as required.
    \end{proof}
    
\end{theorem}

\begin{remark}
    This result shows that timing correlations are approximately preserved under dynamics that nearly commute with the synchronization operator. The deviation from perfect timing alignment grows linearly with time, with slope proportional to $\epsilon$. This behavior formalizes a notion of robustness for time-encoded quantum states under small Hamiltonian imperfections.
\end{remark}

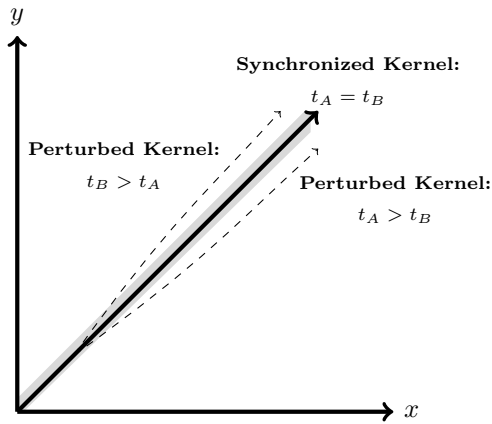
\begin{figure}[h]
\centering 
%\tdplotsetmaincoords{60}{120}%
%\tdplotsetrotatedcoords{0}{20}{0}%
\begin{tikzpicture}
%\draw[help lines, color=gray!30, solid]  grid (5.0,5.0);

\draw[->,ultra thick] (0,0)--(5,0) node[right]{$x$};
\draw[->,ultra thick] (0,0)--(0,5) node[above]{$y$};
% Parameters
\def\epsilon{0.18}  % Narrow band

% Shaded epsilon band around diagonal
%\fill[blue!10, opacity=0.75]
%  (0.02, \epsilon) -- (3.9, 3.9+\epsilon) -- (3.9, 3.9-\epsilon) -- (0.02, -\epsilon) -- cycle;
% Shaded epsilon band around diagonal (safe and clean)
\begin{scope}
  \clip (0,0) rectangle (4.0,4.0); % Keep shading in the first quadrant
  \fill[black, opacity=0.15]
    (0.0, 0.0+\epsilon) -- (3.9, 3.9+\epsilon) -- (3.9, 3.9-\epsilon) -- (0.0, 0.1-\epsilon) -- cycle;
\end{scope}

\draw[->, ultra thick] (0,0)--(4,4) node[pos=1.1, align=center]
{
\scriptsize \textbf{Synchronized Kernel:}\\
\scriptsize $t_A = t_B$
};

%\filldraw [gray] (1.5, 1.5) circle [radius=2pt]
 %                (2.5, 2.0) circle [radius=2pt]
 %                (2.9, 2.5) circle [radius=2pt]
 %                (3.1, 3.5) circle [radius=2pt]
 %                
 %                (2.0, 2.5) circle [radius=2pt]
 %                (2.5, 2.9) circle [radius=2pt]
 %                (3.5, 3.1) circle [radius=2pt];
%\draw[->, dashed] (1.5, 1.5) .. controls (2.2, 2.2) .. (3.5, 3.5);
\draw[->, dashed, black] (0.75, 0.75) .. controls (2.5, 2.0) and (2.9, 2.5) .. (4.0, 3.5) 
node[pos=0.75, right, xshift=10pt, align=center]{\scriptsize \textbf{Perturbed Kernel:}\\ \scriptsize $t_A > t_B$};

\draw[->, dashed, black] (0.75, 0.75) .. controls (2.0, 2.5) and (2.5, 2.9) .. (3.5, 4.0) 
node[left, pos=0.75, align=center]{\scriptsize \textbf{Perturbed Kernel:}\\ \scriptsize $t_B > t_A$};
  
\end{tikzpicture}
\caption{
Visualization of synchronization dynamics under perturbative evolution. The solid diagonal line represents the ideal synchronization condition \( t_A = t_B \). The shaded band of width \( 2\epsilon \) illustrates a tolerance window, capturing approximate synchronization as defined by the kernel-preserving bounds in Theorem~6.2. Dashed lines show trajectories under \( \epsilon \)-compatible Hamiltonians, which deviate from the diagonal but remain confined within the acceptable synchronization region.
}

\end{figure}

\begin{corollary}[Stability Over Short Time Intervals]
Let $\ket{\psi(0)} \in \ker(T_A \otimes I - I \otimes T_B)$, and $H$ be $\epsilon$-compatible. Then for any $\delta > 0$, the evolved state satisfies:

\begin{equation}
\|(T_A \otimes I - I \otimes T_B)\ket{\psi(t)}\| \le \delta \quad \text{for all } |t| \le \frac{\delta}{\varepsilon}
\end{equation}
This shows that perfect timing correlations are approximately preserved over time intervals $|t| \lesssim \frac{\delta}{\varepsilon}$, with deviation growing in linear time.

\end{corollary}

\begin{corollary}[Fidelity to the Kernel Subspace]
Let $\mathcal{K} := \ker(T_A \otimes I - I \otimes T_B)$ and let $\Pi_\mathcal{K}$ be the orthogonal projection onto $\mathcal{K}$. For $\epsilon$-compatible Hamiltonians $H$, the fidelity of the evolved state satisfies:
\begin{equation}
    F(t) := ||\Pi_\mathcal{K} \ket{\psi(t)}||^2 \geq 1 - \epsilon^2 t^2
\end{equation}
    This bound provides a second-order estimate on the degradation of synchronization fidelity over time.
\end{corollary}

The results in this section demonstrate that the time-correlation structure encoded in the kernel of $T_A \otimes I - I \otimes T_B$ is robust under small perturbations to the dynamics. When the Hamiltonians are $\epsilon$-compatible with the clock observables, deviations from perfect synchronization grow linearly in time and remain negligible over short intervals. This suggests that even in imperfect physical systems, quantum timing information can remain coherent over useful time windows provided the evolution remains close to the ideal commuting model.

\begin{question}[Long-Time Stability and Error Correction]
Can the timing-correlation structure be actively preserved or recovered over longer time scales, particularly in systems affected by noise, decoherence, or environmental drift? In practical regimes such as free-space quantum communication, perturbations may exceed the $\epsilon$-compatible range. Could timing fidelity be extended using error-correcting codes, feedback control, or engineered symmetries? More generally, are there structural conditions under which the deviation from the kernel subspace remains uniformly bounded even as $t \rightarrow \infty$?
\end{question}

\section{Algebraic Classification of Kernel-Preserving Dynamics}

In the previous sections, we analyzed timing correlations in quantum systems through the kernel structure of local time observables. We now formulate a more general and intrinsic classification of synchronization-preserving dynamics using representation theory and operator algebras. In particular, we show that synchronization arises as a structural feature of matched irreducible representations under a group symmetry, and that the class of Hamiltonians preserving timing correlations is characterized by a commutant algebra aligned with this symmetry. This algebraic framework reveals synchronization not merely as an operational phenomenon, but as a spectral invariant within distributed quantum systems.

\begin{theorem}[Algebraic Classification of Synchronization-Preserving Dynamics]
    Let $G$ be a finite group and let $\rho_A :G \rightarrow \mathcal{U(H)_A}$ and $\rho_B : G \rightarrow \mathcal{U(H)}_B$ be unitary representations on finite-dimensional Hilbert spaces. Define the joint representation
    \begin{equation}
        \rho(g) := \rho_A(g) \otimes \rho_B(g),\ \text{acting on } \mathcal{H} := \mathcal{H}_A \otimes \mathcal{H}_B.
    \end{equation}

    Let $\mathcal{K}_G \subseteq \mathcal{H}$ be the subspace:
    \begin{equation}
        \mathcal{K}_G := \bigoplus_{\lambda \in \hat{G}} (V_\lambda \otimes V_\lambda),
    \end{equation}
    where $\hat{G}$ denote the set of isomorphism classes of irreducible representations \textit{irreps} of the group $G$, denotes the set of irreducible representations of $G$, and the sum is taken over the irrep types in the decomposition of $\mathcal{H}_A$ and $\mathcal{H}_B$. Then $\mathcal{K}_G$ is invariant under the joint action $\rho(g)$ for all $g \in G$, and forms the diagonal isotypic subspace with respect to $G$.  Let $T_A \in \text{End}_G(\mathcal{H}_A)$ and $T_B \in \text{End}_G(\mathcal{H}_B)$ be self-adjoint operators that commute with the group action:

        \begin{equation}
            [\rho_A(g), T_A] = 0,\ [\rho_B(g), T_B] = 0\ \  \forall g \in G
        \end{equation}
         Define the synchronization kernel operator:
        \begin{equation}
            K := T_A \otimes I - I \otimes T_B.
        \end{equation}
        
        Then $K$ commutes with the joint action $\rho(g)$, $K_G \subseteq \ker(K)$, and any $H \in \text{End}_G(\mathcal{H}$ satisfying $[H, K] = 0$ preserves $K_G$. If $T_A, T_B \in Z(\mathbb{C}[G])$, i.e. central  elements of the group algebra, then the spectral projections of $T_A$ and $T_B$ align with the isotypic decomposition of $\mathcal{H}_A, \mathcal{H}_B$, and the kernel condition
        \begin{equation}
            (T_A \otimes I - I \otimes T_B)\ket{\psi} = 0
        \end{equation}
        is equivalent to synchronization of irrep labels.

    \begin{proof}

        % 1st Step
        The space $\mathcal{K}_G$ is a direct sum over $\lambda \in \hat{G}$ of subspaces of the form $V_\lambda \otimes V_\lambda$. Each subspace is invarient under the joint action:
        \begin{equation}
            \rho(g) = \rho_A(g) \otimes \rho_B(g)
        \end{equation}
        Since both factors are irreducible representations of the same type $\lambda$, their tensor product is invariant under the group action:

        \begin{equation}
            \rho (g)(v \otimes w) = \rho_A (g)v \otimes \rho_B (g)w \in V_\lambda \otimes V_\lambda
        \end{equation}
        Therefore, $\mathcal{K}_G$ is $\rho(G)$-invariant. \\ 

        %2nd Step
        Let $T_A \in \operatorname{End}_G(\mathcal{H}_A)$, so $[T_A, \rho_A(g)] = 0$ for all $g \in G$, and similarly for $T_B$. Then:
        \begin{align*}
            [\rho(g), T_A \otimes I] = [\rho_A(g) \otimes \rho_B(g), T_A \otimes I] \\= [\rho_A(g), T_A] \otimes \rho_B(g) = 0
        \end{align*}
        
        Likewise,
        \begin{equation}
            [\rho(g), I \otimes T_B] = \rho_A(g) \otimes [\rho_B(g), T_B] = 0
        \end{equation}
        Thus, $[\rho(g), K] = 0$ for all $g \in G$.\\

        To show $\mathcal{K}_G \subseteq \ker(K)$, each $V_\lambda$ carries and irreducible representation and since $T_A \in  \operatorname{End}_G(\mathcal{H}_A)$, by Schur's Lemma \cite{schur-duality}, the restriction of $T_A$ to each irreducible subspace is proportaionl to the identity:

        \begin{equation}
            T_A \mid_{V_{\lambda}} = \alpha_\lambda  I,\ T_B \mid_{V_{\lambda}} = \beta_\lambda  I
        \end{equation}
        Thus, for any $v \in V_\lambda \otimes V_\lambda \subset \mathcal{K}_G$, we compute:
        \begin{equation}
            K(v \otimes w) = (\alpha_\lambda - \beta_\lambda)v \otimes w.
        \end{equation}
        If $\alpha_\lambda = \beta_\lambda$, then $v \otimes w \in \ker(K)$. If not, $v \otimes w \notin \ker(K)$, but still lies in an isotypic summand.
        Henceforth, if $T_A$ and $T_B$ assign the same scalar to each $V_\lambda$, then $\mathcal{K}_G \subseteq \ker(K)$. In particular, this  always holds when $T_A = T_B$ or both are derived from a common central element of $Z(\mathbb{C}[G])$ which will be subsequently addressed.\\

        %3rd Step 
        Let $H \in \operatorname{End}_G(\mathcal{H})$, i.e., $H$ commutes with the joint representation $\rho(G)$, and suppose $[H,K] = 0$. Then both $H$ and $K$ lie in the commutant $\rho(G)'$, which is known to be block-diagonal over the decomposition into irreducible:
        \begin{equation}
            \mathcal{H} \cong \bigoplus_{\lambda,\mu}(V_\lambda \otimes V_\lambda) \otimes \mathbb{C}^ {m_{\lambda,\mu} }
        \end{equation}
        So each operator preserves the isotypic components. Since $H$ and $K$ commute and preserve these components, and $K$ annihilates $\mathcal{K}_G$, then $H$ must map $\mathcal{K}_G$ into itself. Hence, $H\mathcal{K}_G \subseteq \mathcal{K}_G$, i.e., $H$ preserves the synchronization subspace. 

        %4th step
        Suppose $T_A, T_B \in Z(\mathbb{C}[G])$, the center of the group algebra. Then they act as a class functions and by Schur's Lemma \cite{schur-duality}, act as scalar multiples of the identity on each irreducible representation:

        \begin{equation}
        T_A \big|_{V_\lambda} = \gamma_\lambda I, \qquad
        T_B \big|_{V_\lambda} = \gamma_\lambda I
        \end{equation}
        
        %\begin{equation}
        %    T_A\mid_{V_\lambda} = \gamma_\lambda I = T_B \mid_{V_\lambda}
        %\end{equation}

        Thus, for each matched $V_\lambda \otimes V_\lambda \in \mathcal{K}_G$, the operator 
        \begin{equation}
            K = T_A \otimes I - I \otimes T_B
        \end{equation}

        acts as zero. Hence:
        \begin{equation}
            ker(K) = \bigoplus_{\lambda :_{\gamma_\lambda = \gamma_\lambda}} V_\lambda \otimes V_\lambda = \mathcal{K}_G
        \end{equation}
        proving that synchronization corresponds exactly to spectral alignment under the representation levels.

    \end{proof}

\end{theorem}

\begin{corollary}[Algebra of Synchronization-Preserving Dynamics]
    Let $G$ be a finite group with unitary representations $\rho_A : G \rightarrow \mathcal{U(H}_A), \rho_B : G \rightarrow \mathcal{U(H}_B)$ and let $\rho(g) := \rho_A(g) \otimes \rho_B(g)$ act on $\mathcal{H} := \mathcal{H}_A \otimes \mathcal{H}_B$. Let $T_A \in \operatorname{End}_G(\mathcal{H}_A), T_B \in \operatorname{End}_G(\mathcal{H}_B)$ be self-adjoint time observables and define: 
    \begin{equation}
        K := T_A \otimes I - I \otimes T_B,\ \mathcal{K} := \ker(K).
    \end{equation}
    Then the set $\mathcal{H}_{sync} := { H \in \operatorname{End}_G(\mathcal{H} \mid [H, K] = 0}$ forms a unital *-subalgebra of $\mathcal{B(H)}$ that preserves the synchronization subspace $\mathcal{K}$. $\mathcal{H}_{sync}$ is the maximal algebra of $G$-equivalent Hamiltonians that preserve synchronization encoded by $K$. That is, any operator outside $\mathcal{H}_{sync}$ necessarily evolves some synchronized state outside of $\mathcal{K}$. If $T_A = T_B \in Z(\mathbb{C}[G])$, then:
    \begin{equation}
        \mathcal{K} := \mathcal{K}_G := \bigoplus_{\lambda \in \hat{G}} V_\lambda \otimes V_\lambda
    \end{equation}
    and $\mathcal{H}_{sync}$ is the commutant of the joint action and the kernel operator:
    \begin{equation}
        \mathcal{H}_{sync} = \{H \in \rho(G)' \mid [H,K] = 0\}
    \end{equation}

\end{corollary}

\begin{remark}
    This shows that synchronization is not preserved merely by accident or weak commutation, but it is fully characterized by a symmetry-respecting operator algebra. The condition $[H, K] = 0$ selects precisely those dynamics that leave timing correlations structurally intact. 
\end{remark}

\begin{example}[Pauli-Z Qubits and Synchronization Preservation]
    We illustrate the group-theoretic classification of synchronization-preserving dynamics with a simple bipartite system of two qubits. Let $\mathcal{H}_A = \mathcal{H}_B = \mathbb{C}^2$ and define the local clock observable as
    \begin{equation}
        T_A = T_B = \sigma_z = \begin{pmatrix}
            1 & 0 \\
            0 & -1 
        \end{pmatrix}
    \end{equation}
    where $\sigma_z$ is the standard Pauli-Z matrix. The synchronization observable is then given by
    \begin{equation}
        K = T_A \otimes I - I \otimes T_B.
    \end{equation}
    The synchronized subspace $\mathcal{S} \subset \mathcal{H}_A \otimes \mathcal{H}_B$ is the kernel of $K$, consisting of all bipratite states $\ket{\psi}$ satisfying $S \ket{\psi} = 0$. A straightforward computation shows that $\mathcal{K}$ is spanned by the states $\ket{00}$ and $\ket{11}$, where $\ket{0}$ and $\ket{1}$ are eigenstates of $\sigma_z$ with the eigenvalues $1$ and $-1$, respectively. To preserve synchronization, the system must evolve under Hamiltonians that commute with $S$. Since $T_A$ and $T_B$ are diagonal in the computational basis, any Hamiltonian $H$ of the form
    \begin{equation}
        H = a(Z \otimes I) + b(I \otimes Z),
    \end{equation}
    with real coefficients $a, b \in \mathbb{R}$, satisfies $[H,K] = 0$. More generally, any Hamiltonian diagonal in the computational basis, possibly including terms proportional to $Z \otimes Z$, also preserves the synchronization. 

    The set of synchronization-preserving Hamiltonians forms a commutative *-algebra generated by $Z \otimes I$ and $I \otimes Z$. Group-theoretically, this corresponds to the trivial representation of the abelian group $\mathbb{Z}_2 \times \mathbb{Z}_2$, where each generator acts diagonally on the computational basis states. The synchronized subspace $\mathcal{K}$ corresponds to a two-dimensional invariant subspace under this representation. 

    This simple two-qubit example illustrates the key features of our classification: the preservation of synchronization can be understood in terms of invariant subspaces under the action of a commutative *-algebra generated by commuting observables. In larger or more complex systems, the synchronization subspace similarly corresponds to a collection of irreducible representations matched across subsystems. This structural viewpoint not only clarifies the conditions for synchronization stability but also suggests natural extensions to systems with richer symmetry groups or higher-dimensional clock observables, as discussed in Section 8.
\end{example}

The classification of synchronization-preserving dynamics developed above highlights that timing coherence is not an accidental property of particular observables, but rather an emergent feature of underlying symmetry and algebraic structure. Synchronization corresponds to spectral alignment across matched irreducible components, and its preservation is governed by the algebra of symmetry-respecting, kernel-commuting operators. This perspective suggests several natural generalizations, including extensions to multipartite systems, categorical frameworks, and timing protocols constrained by group symmetries, which we outline in the following section.

\section{Generalizations and Open Directions}

This work has developed a structural and algebraic framework for modeling quantum time transfer via observable-based synchronization. By identifying the synchronization subspace as the kernel of commuting observables and classifying its preservation under compatible dynamics, we have constructed a minimal operator-theoretic foundation for timing correlations in bipartite quantum systems. Several directions remain open for future exploration:

\begin{enumerate}
    \item \textbf{Multipartite Extensions.} The kernel structure can be generalized to multipartite systems by considering simultaneous preservation of multiple pairwise or global observable alignments. Given observables $T_1, T_2, \dots, T_n$, one may define a synchronized subspace through intersections of pairwise kernels or higher-order symmetry conditions. The structure and dynamics of such subspaces in networked or topologically constrained systems merit further classification.

    \item \textbf{Timing as a Resource.} The fragility of synchronization under generic evolution suggests a possible resource-theoretic formulation. One could ask whether timing correlations can be quantified, preserved, or distilled under constrained operations, analogously to entanglement and coherence \cite{coecke2016resources}.
    
    \item \textbf{Connections to Error Correction and Memory.} The algebraic constraints identified in Section 7 resemble stabilizer conditions in quantum error correction. This invites the question of whether synchronization subspaces can be protected, corrected, or stabilized through active coding schemes, potentially enabling long-term timing fidelity in realistic noisy environments.

    \item \textbf{Beyond Diagonal Observables.} While this paper focused on diagonal clock observables, one could consider timing observables with degenerate, non-orthogonal, or more generally normal (non-diagonalizable) spectra. This would require reformulating synchronization in terms of spectral projections, commutants, or higher categorical structures.
\end{enumerate}

These generalizations open the door to a broader mathematical understanding of quantum timing structure, symmetry, and resource theories beyond the bipartite, diagonal setting considered here. The results establish a foundation for the systematic study of quantum synchronization as a structural and algebraic phenomenon, bridging timing, symmetry, and information in distributed quantum systems. At a structural level, the preservation of synchronization subspaces under compatible evolution suggests a potential categorical generalization. Synchronization-preserving unitaries may be viewed as morphisms in a category whose objects are observables equipped with compatible dynamics, and whose commuting diagrams express subspace invariance. Developing this perspective could offer a unifying abstraction across operator theory, quantum information, and categorical quantum mechanics.

\appendix
\section{Group-Theoretic Structures and \\ Symmetry in Timing Observables}

The algebraic framework developed in the main body of this paper invites generalization to settings where observables and Hamiltonians exhibit global symmetries, potentially arising from finite or Lie group actions. We outline a representation-theoretic perspective on time observables and synchronization-preserving dynamics and suggest a pathway to future classification results based on symmetry groups.

\subsection{Representation Theory of Finite Groups}

\begin{proposition}[Unitary Decomposition into Irreducibles]
Let $G$ be a finite group, and let $\rho : G \rightarrow \mathcal{U(H)}$ be a finite-dimensional unitary representation. Then the Hilbert space $\mathcal{H}$ decomposes as a direct sum:
\begin{equation}
    \mathcal{H} \cong \bigoplus_{\lambda \in \hat{G}} V_\lambda \otimes \mathbb{C}^{m_{\lambda}}
\end{equation}
where $\hat{G}$ denotes the irreducible representations of $G$, $V_\lambda$ is a representative irreducible space for each $\lambda \in \hat{G}$, and $m_\lambda \in \mathbb{N}_0$ is the multiplicity of $V_\lambda \in \mathcal{H}$ for which the decomposition is unique up to unitary equivalence. 

\begin{proof}
    Since $G$ is finite and $\rho$ is unitary, by Maschke's Theorem \cite{fulton1991representation, serre1977linear}, the representation of $\rho$ is completely reducible: that is, $\mathcal{H}$ can be decomposed into an orthogonal direct sum of irreducible invariant subspaces. Furthermore, since the group algebra $\mathbb{C}[G]$ is semi-simple, the regular representation decomposes into a direct sum of all irreducible representations, with multiplicities corresponding to their dimensions. Let $\{V_\lambda\}_{\lambda \in \hat{G}}$ denote a full set of pairwise non-isomorphic irreducible unitary representations of $G$. By standard theory \cite{fulton1991representation, serre1977linear}, we have
    \begin{equation}
        \mathcal{H} \cong \bigoplus_{\lambda \in \hat{G}} V_\lambda \otimes \mathbb{C}^{m_{\lambda}},
    \end{equation}
    where $m_\lambda$ counts the multiplicity of $V_\lambda \in \mathcal{H}$. Each $V_\lambda \otimes \mathbb{C}^{m_{\lambda}}$ can be thought of as $m_\lambda$ independent copies of $V_\lambda$. Uniqueness up to unitary equivalence follows from the complete reducibility and the orthogonality relations for irreducible characters. 
\end{proof}
    
\end{proposition}

\subsection{Schur's Lemma and Equivalent Maps}

\begin{theorem}[Schur's Lemma for Finite Groups]
Let $G$ be a finite group, and let $V, W$ be finite-dimensional irreducible unitary representations of $G$ over $\mathbb{C}$. Suppose $A : V \rightarrow W$ is a linear map satisfying
\begin{equation}
    A\rho v(g) = \rho w(g)A,\ \forall g \in G,
\end{equation}
where $\rho v$ and $\rho w$ denote the respective group actions. If $V \cong W$, $V$ and $W$ are isomorphic as representations and $A$ is a scalar multiple of the identity operator on $V$. If $V \not\cong W$ then $A = 0$ \cite{fulton1991representation}.

\begin{proof}
    Assume $A \neq 0$, since $A$ intertwines the actions of $G$, we observe that the kernel $\ker(A) \subseteq V$ is invariant under the action of $G$ and the image $\operatorname{Im}(A) \subseteq W$ is also invariant under the action of $G$. For any $g \in G$ and $v \in \ker(A)$, we have:
    \begin{equation}
        A(\rho v(g)v) = \rho w(g)A(v) = \rho w(g)(0) = 0,
    \end{equation}

    so $\rho v(g) \in \ker(A)$. Similarly, for any $g \in G$ and $w = Av \in \operatorname{Im}(A)$,
    \begin{equation}
        \rho w(g)w = \rho w(g)A(v) = A(\rho v(g) v) \in \operatorname{Im}(A).
    \end{equation}
    Thus, $\ker(A)$ is a $G$-invariant subspace of V, and $\operatorname{Im}(A)$ is a $G$-invariant subspace of $W$. Since $V$ and $W$ are irreducible representations, their only invariant subspaces are $\{0\}$ and the entire space. Therefore the following two cases arise:
    \begin{enumerate}
        \item $\ker(A) = V$, then $A = 0$ contradicting $A \neq 0$ hence $\ker(A) = \{0\}$, so $A$ is injective.

        \item $\operatorname{Im}(A) = 0$, again $A = 0$ and once again contradicting $A \neq 0$, hence $\operatorname{Im}(A) = W$ so A is surjective.
    \end{enumerate}
    We can then conclude $A$ is a bijection and an isomorphism to the vector spaces.

    Now since $A$ intertwines the group actions and is invertible, it follows that $V \cong W$ as representations, furthermore when $V = W$, by Schur's Lemma \cite{schur-duality}, in the classical form, the space of interwining operators $\hom_G(V,V)$ is one-dimensional, consisting of scalar multiples of the identity. Therefore, $A = \lambda I$ for some $\lambda \in \mathbb{C}$

    In the case $V \not\cong W$, there can be no non-zero intertwining operator between non-isomorphic irreducibles. 
    
\end{proof}

\end{theorem}

\subsection{Tensor Products and Synchronization Subspaces}

\begin{proposition}[Diagonal Synchronization Subspace via Irreducibles]
    Let $G$ be a finite group and let $\rho_A : G \rightarrow \mathcal{U(H}_A)$ and $\rho_B : G \rightarrow \mathcal{U(H}_B)$ We define the joint action:
    \begin{equation}
        \rho(g) := \rho_A(g) \otimes \rho_B(g)\ \text{on}\ \mathcal{H} := \mathcal{H}_A \otimes \mathcal{H}_B.
    \end{equation}

    Then there exists a distinguished subspace 
    \begin{equation}
        \mathcal{K}_G := \bigoplus_{\lambda \in \hat{G}} V_\lambda \otimes V_\lambda,
    \end{equation}

    where $V_\lambda$ runs over the irreducible representations of $G$ appearing simultaneously in both $\mathcal{H}_A$ and $\mathcal{H}_B$, such that $\mathcal{K}_G$ is invariant under the joint action $\rho(G)$ and $\mathcal{K}_G$ consists precisely of states synchronized across irreducible representation types.

    \begin{proof}
        Since $G$ is finite and $\rho_A$ and $\rho_B$ are unitary, such representations decomposes as:
        \begin{equation}
            \mathcal{H}_A \cong \bigoplus_{\lambda \in \hat{G}} V_\lambda \otimes \mathbb{C}^{m_{\lambda}},\ \mathcal{H}_B \cong \bigoplus_{\lambda \in \hat{G}} V_\lambda \otimes \mathbb{C}^{m_{\lambda}},
        \end{equation}
        where $m_\lambda, n_\mu \geq 0$ are multiplicities. Then the tensor product space $\mathcal{H} = \mathcal{H}_A \otimes \mathcal{H}_B$ decomposes into sectors labeled by pairs $(\lambda, \mu)$. The sector $V_\lambda \otimes V_\mu$ transforms under the joint action $\rho(g) = \rho_A(g) \otimes \rho_B(g)$ as:
        \begin{equation}
            \rho(g)(v \otimes w) = \rho_A(g)v \otimes \rho_B(g)w.
        \end{equation}

        When $\lambda = \mu$, the two factors transform according to the same irreducible representations. Thus, the direct sum
        
        \begin{equation}
            \mathcal{K}_G := \bigoplus_{\lambda \in \hat{G}} V_\lambda \otimes V_\lambda
        \end{equation}
        consists precisely of states where the representation type matches on both sides. Moreover, $\mathcal{K}_G$ is invariant under $\rho(G)$ because the action preserves the tensor structures of each $V_\lambda \otimes V_\lambda$.

    \end{proof}
\end{proposition}

\subsection{Kernel of Time Observables and Synchronization}

\begin{proposition}[Synchronization via Time Observables and Kernel Structure]
    let $G$ be a finite group and let $\rho_A : G \rightarrow \mathcal{U(H}_A)$ and $\rho_B : G \rightarrow \mathcal{U(H}_B)$ be finite-dimensional unitary representations. Suppose $T_A \in \operatorname{End}_G(\mathcal{H}_A)$ and $T_B \in \operatorname{End}_G(\mathcal{H}_B)$ are self-adjoint operators commuting with the respective group actions. Define the synchronization operator 
    \begin{equation}
        K := T_A \otimes I - I \otimes T_B\ \text{on}\ \mathcal{H} := \mathcal{H}_A \otimes \mathcal{H}_B.
    \end{equation}
    Then the operator $K$ commutes with the joint action $\rho(g) := \rho_A(g) \otimes \rho_B(g)\ \forall g \in G$. The synchronized subspace
    \begin{equation}
        \mathcal{K}_G := \bigoplus_{\lambda \in \hat{G}} V_\lambda \otimes V_\lambda
    \end{equation}
    (as defined in Proposition A.3) satisfies $\mathcal{K}_G \subseteq \ker(K)$.
    
    \begin{proof}
        Since $T_A \in \operatorname{End}_G(\mathcal{H}_A)$ and $T_B \in \operatorname{End}_G(\mathcal{H}_B)$, by definition we have
        \begin{equation}
            \rho_A(g)T_A = T_A\rho_A(g),\  \rho_B(g)T_B = T_B\rho_B(g), \forall g \in G.
        \end{equation}
        Thus:
        \begin{equation}
            \rho(g)(T_A \otimes I) = (\rho_A(g)T_A) \otimes \rho_B(g) = (T_A\rho_A(g)) \otimes \rho_B(g) = (T_A \otimes I)\rho(g),
        \end{equation}
        and similarly
        \begin{equation}
            \rho(g)(I \otimes T_B) = (I \otimes T_B)\rho(g).
        \end{equation}
        Therefore, $K$ commutes with $\rho(g)$ for all $g \in G$.

        Next to prove $\mathcal{K}_G \subseteq \ker(K)$, recall from Proposition A.3 that $\mathcal{K}_G$ consists of states $v \otimes w$ where $v,\ w \in V_\lambda$ for the same $\operatorname{irrep} \lambda$. By Schur's Lemma \cite{schur-duality} and from Theorem A.2, since $T_A$ and $T_B$ are $G$-equivariant, each acts as a scalar multiple of the identity on $V_\lambda$:
        \begin{equation}
            T_{A\mid V_{\lambda}} = t_\lambda I,\ T_{B\mid V_{\lambda}} = t_\lambda I,
        \end{equation}
        for some $t_\lambda \in \mathbb{R}$. Thus, for any $v \otimes w \in V_\lambda \otimes V_\lambda$, we compute:
        \begin{equation}
            (T_A \otimes I)(v \otimes w) = T_{A}v \otimes w = t_\lambda v \otimes w,\ (I \otimes T_B)(v \otimes w) = v \otimes T_Bw = t_\lambda v \otimes w
        \end{equation}
        and hence
        \begin{equation}
            K(v \otimes w) = (T_A \otimes I - I \otimes T_B)(v \otimes w) = 0.
        \end{equation}
        Therefore, $v \otimes w \in \ker(K)$, and thus $\mathcal{K}_G \subseteq \ker(K)$

    \end{proof}

\end{proposition}

\subsection*{ Outlook and Structural Extensions}

The results above highlight that synchronization-preserving dynamics are intimately tied to group-theoretic and representation-theoretic structure. In particular, Theorems A.1–A.4 show how symmetric time observables and alignment subspaces arise naturally from group actions and commutative operator algebras. These findings invite a number of structural generalizations.

One direction involves classifying synchronization-preserving Hamiltonians when the observables $T_A$  and $T_B$ are constructed from central elements of a group algebra, or lie in the commutant of a common symmetry representation. In such settings, timing correlations may be interpreted through the decomposition of tensor product representations, with synchronization corresponding to the alignment of irreducible components.

This framework suggests deeper connections to the representation theory of symmetric groups $S_n$, wreath products, and their associated centralizer algebras, which are well-suited for modeling multipartite or permutation-invariant timing structures. More broadly, a categorical formulation—treating observables as functors or module homomorphisms between representation categories—may offer a natural language for encoding quantum time and synchronization in highly structured or networked systems.

These directions position group symmetry and its representation-theoretic structure as a promising foundation for extending quantum time transfer beyond the setting of diagonal commuting observables, toward a broader class of symmetry-protected timing protocols.

\section*{Acknowledgments}
The author thanks Dr. R. Nicholas Lanning and Aaron Marcus for detailed feedback on earlier drafts of this paper and helpful discussions. 

\bibliographystyle{amsplain}
\bibliography{references}

\end{document}